\documentclass[preprint,showpacs,amsfonts,amsmath,amssymb]{revtex4}
\usepackage{graphicx}% Include figure files
\usepackage{dcolumn}% Align table columns on decimal point
\usepackage{bm}% bold math
\usepackage{showlabels}% remove in final version

\begin{document}
%\preprint{example}
\title{ Evolution of the interacting viscous dark energy model in Einstein cosmology  }
%\altaffiliation{}
\author{Juhua Chen$^{1,2}$} \email{jhchen@hunnu.edu.cn}
\author{Yongjiu Wang$^{1}$}
\affiliation{College of Physics and Information Science, Hunan
Normal University, Changsha, Hunan 410081, P. R. China
\\Department of Physics \& Astronomy,
      University of Missouri, Columbia, MO 65211, USA}
\begin{abstract}

In this paper we investigate the evolution of  the viscous  dark
energy (DE) interacting with the dark matter (DM) in the Einstein
cosmology model. Using the linearizing theory of the dynamical
system, we find, in our model, there exists a stable late time
scaling solution which corresponds to the accelerating universe, and
we also find the unstable solution under some appropriate
parameters. In order to alleviate the coincidence problem, some
authors considered the effect of quantum correction due to the
conform anomaly and the interacting dark energy model. But if we
take into account the bulk viscosity of the cosmic fluid, the
coincidence problem will be softened just like the interacting dark
energy cosmology model. That's to say, both the non-perfect fluid
model and the interacting models of the dark energy can alleviate or
soften the singularity of the universe.


\end{abstract}

\maketitle

\section{Introduction}
It's well known that recent data from Ia supernova (SN Ia)
\cite{Pelmutter} and microwave background (CMB) radiation
\cite{Bennett} have provided strong evidences for a spatially flat
and accelerating universe in the present time. The origin of
accelerating expansion is regarded that the universe is dominated by
an exotic component with the negative pressure called "dark energy"
which constitutes 70 percent of the energy density of the universe,
and dark matter about 26 percent.  There are several candidates for
dark energy: The first is the cosmological constant
\cite{Padmanabham}, and the second is the so-called dynamic
candidates such as: Phantom \cite{Caldwelland}, quintessence
\cite{Capozziello}, K-essence \cite{Chiba} and quintom \cite{Wei}.
The difference of these candidates for dark energy is the size of
the parameter $\omega_{E}$, namely the ration of the pressure and
energy density of the dark energy. For quintessence, the state
equation is given by the relation between the pressure $p_{E}$ and
the energy density $\rho_{E}$, i.e. $ p_{E}=\omega_{E}\rho_{E}$,
where $ -1<\omega_{E}<-1/3$. The borderline case of  $
\omega_{E}=-1$ of the extraordinary quintessence covers the
cosmological constant term.

The negative pressure of the dark energy may be the cause of the
acceleration of the present Universe. However, the nature of the
dark energy still remains a complete mystery. No more than five
years ago, some physicists \cite{Caldwell} found that, if we only
assumed the cosmic fluid to be ideal, i.e. nonviscous, it must bring
out the occurrence of a singularity of the universe in the far
future. There are two methods to modify or soften the singularity.
The first is the effect of quantum corrections due to the conformal
anomaly \cite{Brevik}. The other is to consider the bulk viscosity
of the cosmic fluid \cite{Misner}. The viscosity theory of
relativistic fluids was first suggested by Eckart, Landau and
Lifshitz \cite{Eckart}. In recent years some physicists
\cite{Belinsky} also took into account  the bulk viscous cosmology.
In this paper we consider that the bulk viscous dark energy is
characterized by energy density $\rho_{E}$ and pressure $p_{E}$ as
$p_{E}=\omega_{E}\rho_{E}+\beta\rho_{E}^{d}$.

Up to now the model, which the dark energy  is considered as the
perfect fluid state, suffers the coincidence problem
\cite{Steinhardt}. That's to say, why the dark energy (DE) and the
dark matter (DM) are comparable in size exactly right now. In recent
years some interacting DE models have been brought out to overcome
the problem. In these models they all are assumed that there exists
a nonzero interaction between DE and DM in the Universe and gauges
DE transfers to DM which allows us to create an equilibrium balance
in the evolution of the Universe, so that the density of DE keeps
the same order as that of DM at late times. Some authors \cite{wu}
investigated dynamical behaviors of the dark energy models with only
the dark energy linear equation of state interacting with dark
matter in different cosmology models. They all found that the
universe will enter an era dominated by dark energy and dark matter
with interaction between them, and accelerate in late time under
some proper parameters of the dynamical system. In this paper we
focus on extending the equation of state of dark energy to nonlinear
term, i.e. bulk viscous fluid, to investigate in what way the
nonlinear term affects the evolution of the cosmology.

\section{dynamics of the interacting viscous dark energy model in the flat FRW universe}

Base on the Einstein General Relativity theory, the standard
Friedman equation  acts  as follows in the flat FRW universe
\begin{eqnarray}
H^{2}=\frac{8\pi G}{3}(\rho_{M}+\rho_{E}),
\end{eqnarray}
and the conservation relation of the cosmological total energy is
\begin{eqnarray}
\dot{\rho}+3H(\rho +p)=0,
\end{eqnarray}
where $G$ is the gravitational constant, and the the total energy
density is $\rho=\rho_{M}+\rho_{E}$.

From a hydromechanical standpoint a generalization of cosmic theory
so as to encompass viscosity is most nature. In recent years Misner
et.al \cite{Misner} interestingly investigated the viscous
cosmology. In this paper we consider the general equation of state
equation of the viscous dark energy as
\begin{eqnarray}
p_{E}=\omega_{E}\rho_{E}+\beta\rho_{E}^{d}.
\end{eqnarray}
Under considering the interaction term $Q$ between the dark energy
and dark matter, the evolution equations are
\begin{eqnarray}
\dot{\rho}_{E}+3H[(1+\omega_{E})\rho_{E}+\beta\rho_{E}^{d}]=-Q,\\
\dot{\rho}_{M}+3H\rho_{M}=Q.
\end{eqnarray}

Differentiating Eq.(1), then putting Eq.(2) into it, we can get
\begin{eqnarray}
\dot{H}=-\frac{8\pi G}{2}(\rho_{M}+\rho_{E}+p_{E}).
\end{eqnarray}
If we introduce the follow dimensionless variables
\begin{eqnarray}
x=\frac{8\pi G \rho_{E}}{3H^{2}}, y=\frac{8\pi G
\rho_{M}}{3H^{2}}, \frac{d}{dN}=\frac{1}{H}\frac{d}{dt},
\end{eqnarray}
where $N\equiv lna$ is the number of e-folding to present the
cosmological time. The interacting term  $Q=3bH\rho_{E}$ is between
the dark energy and dark matter. When the the coupling constant $b$
is positive, it means that the dark energy converts into dark
matter. Under these conditions we can construct the follow
autonomous dynamics system by Eqs.(1), (3), (4) and (5) for the
interacting viscous dark energy model in the flat FRW universe.

\begin{eqnarray}
x'=f(x,y)&=&-3(1+b+\omega_{E})x-\beta x^{d}  \nonumber \\
  & &  +x[3y+3(1+\omega_{E})x+\beta x^{d}]\\
y'=g(x,y)&=&3bx-3y+y[3y\nonumber \\ & & +3(1+\omega_{E})x+\beta
x^{d}].
\end{eqnarray}

The dynamics is the general form $X'=F(X)$, where $X$ is the column
matrix constituted by the auxiliary variables and the prime denotes
derivative with respect to $N=lna$. In order to analyze the
stability of the dynamics system, we must linearly expand the
dynamical system near the critical points due to the linearizing
theory of dynamical system. So we can acquire the stability
properties of the dynamical system from the eigenvalues of
linearizing matrix.

In the following section we will focus on evolution of the
dynamical system in the phase space and analyze its stability.

\section{evolution and stability analysis of the autonomous dynamical system}

In order to investigate the evolution  and its stability of the
given dynamical system in the phase space, we take the parameter
$d=2$ with lowest order nonlinear term of the viscous dark energy
$p_{E}$ to its density $\rho_{E}$ for simplicity. We firstly solve
the equations $x'=f(x,y)=0$ and $y'=g(x,y)=0$ to get the critical
points:

 Point I:
 $(x^{I}_{C},y^{I}_{C})=(0,1),$

and Point II:
$(x^{II}_{C},y^{II}_{C})=(\frac{\beta-3\omega_{E}-\gamma}{2\beta},
\frac{\beta+3\omega_{E}+\gamma}{2\beta}),$

where $\gamma=\sqrt{\beta^2+12\beta
b+6\beta\omega_{E}+9\omega_{E}^2}$.

For the Point I, it means that the universe is dominated by the
 dark matter. The eigenvalues of the linearizing matrix for
Point I:
\begin{eqnarray}
\lambda^{I}_{1}=3, \lambda^{I}_{2}=-3(b+\omega_{E}).
\end{eqnarray}

We can see that when $b<-\omega_{E}$, the two eigenvalues are all
positive, so the critical point I is saddle. When $b>-\omega_{E}$,
so the critical point I is node. So we find Point I is an unstable
critical, that's to say, the universe only dominated by the
 dark matter is unstable.

For the Point II,  the existence of the critical point II denotes to
the era  dominated both the dark energy and dark matter in the
late-times of our universe, which is the same as  classical Einstein
cosmology \cite{Olivares}. In Fig.1 we showed the phase diagram of
the autonomous dynamical system in the $(x,y)$ phase space with the
parameters $\beta=0.5,b=0.8,\omega_{E}=-1.2,d=2$.

\begin{figure}[htbp]
\begin{center}
\includegraphics{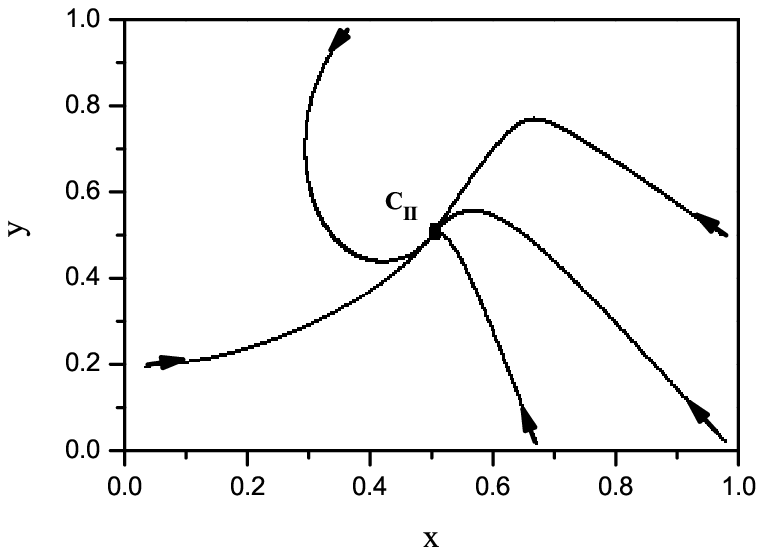}
\caption{The phase diagram of the interacting viscous dark energy
model in the flat FRW universe with the parameters
$\beta=0.5,b=0.8,\omega_{E}=-1.2,d=2$. The point $C_{II}$ is the
critical point of the dynamical system. }
\end{center}
%\end{figure}

%\begin{figure}[htbp]
\begin{center}
\includegraphics{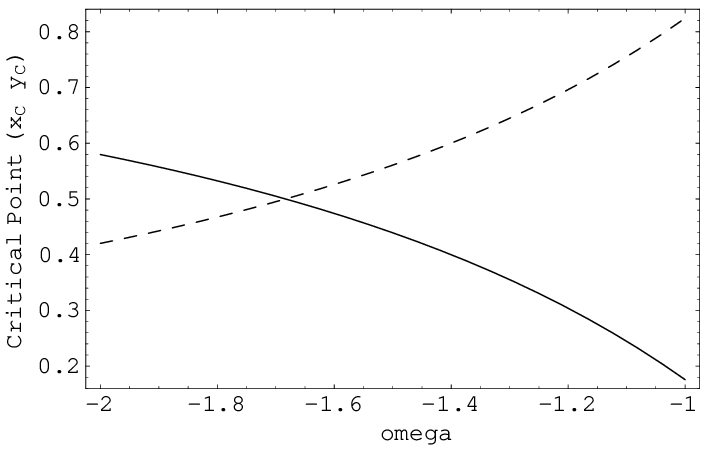}
\caption{Critical point coordinates ($x_{C}$(solid) , $y_{C}$(dash))
for fixed $\beta=0.5, b=0.8$ and $d=2$. }
\end{center}
%\end{figure}
%\begin{figure}[htbp]
\begin{center}
\includegraphics{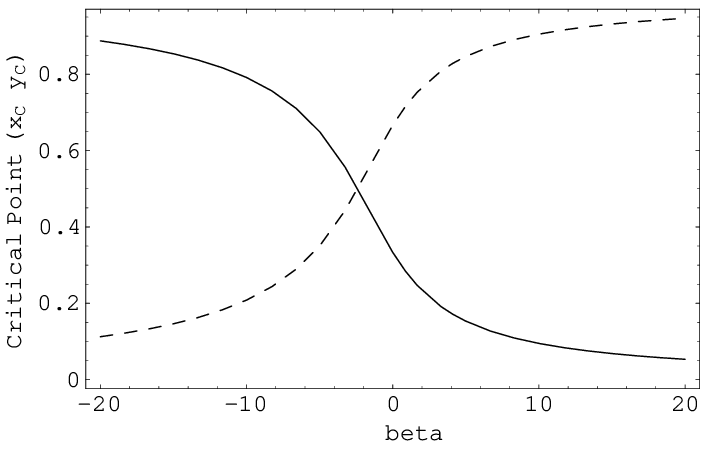}
\caption{Critical point coordinates ($x_{C}$(solid) , $y_{C}$(dash))
for fixed $b=0.8, \omega_{E}=-1.2$ and $d=2$.}
\end{center}
%\end{figure}
%\begin{figure}[htbp]
\begin{center}
\includegraphics{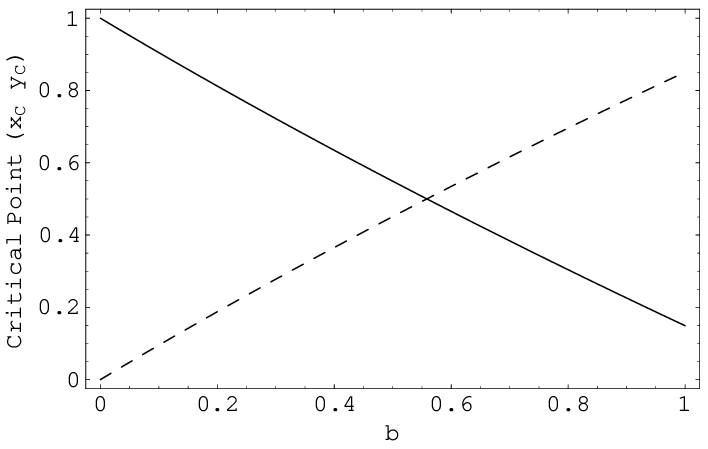}
\caption{Critical point coordinates ($x_{C}$(solid) , $y_{C}$(dash))
for fixed $\beta=0.5, \omega_{E}=-1.2$ and $d=2$.}
\end{center}
\end{figure}

In order to investigate the stability of the critical point II, we
linearize the dynamical system near the critical point II as
follows:
\begin{eqnarray}
\delta x'&=&[-3(b+1+\omega_{E})+3y_{C}+6(1+\omega_{E})x_{C}
\nonumber \\& &-\beta d x_{C}^{d-1}
+\beta(d+1)x_{C}^{d}]\delta x+3x_{C}\delta y,\\
\delta y'&=&[3b+3(1+\omega_{E})y_{C}+\beta d y_{C}x_{C}^{d-1}]\delta
x \nonumber \\& &+ [-3+6y_{C}+3(1+\omega_{E})x_{C}+\beta
x_{C}^{d}]\delta y.
\end{eqnarray}

From the above linearized equations, we can define the following
linearizing matrix
\begin{eqnarray}
M= \left (
         \begin{array}{rr}
        \frac{\partial f}{\partial x} \mid_{x_{C},y_{C}} & \frac{\partial f}{\partial y} \mid_{x_{C},y_{C}} \\
              \frac{\partial g}{\partial x} \mid_{x_{C},y_{C}} &  \frac{\partial g}{\partial y}
              \mid_{x_{C},y_{C}}
          \end{array} \right ),
\end{eqnarray}
where the four elements of the matrix are \\
\begin{eqnarray}
\frac{\partial f}{\partial x}
\mid_{x_{C},y_{C}}&=&-3(b+1+\omega_{E})+3y_{C}+6(1+\omega_{E})x_{C}
\nonumber
\\& &-\beta d x_{C}^{d-1} +\beta(d+1)x_{C}^{d},
\\ \frac{\partial f}{\partial y} \mid_{x_{C},y_{C}}&=& 3x_{C},
\\ \frac{\partial g}{\partial x}\mid_{x_{C},y_{C}}&=&3b+3(1+\omega_{E})y_{C}+\beta d y_{C}x_{C}^{d-1},\\
\frac{\partial g}{\partial y}
\mid_{x_{C},y_{C}}&=&-3+6y_{C}+3(1+\omega_{E})x_{C}+\beta x_{C}^{d}.
\end{eqnarray}

The eigenvalues for the linearizing matrix of the dynamical system
near the critical point II are:
\begin{eqnarray}
\lambda^{II}_{1}&=&\frac{1}{4\beta}(6\beta+2\beta^2+18\beta
b+9\beta\omega_{E}+9\omega_{E}^2-2\beta \gamma
+3\gamma\omega_{E}\nonumber
\\& &-(36\beta^2-72\beta^2b+36\beta^2b^2-36\beta^2\omega_{E} +36\beta^2b\omega_{E}\nonumber \\&
&-108\beta\omega_{E}^2+18\beta^2\omega_{E}^2+216\beta b\omega_{E}^2
+108\beta\omega_{E}^3+162\omega_{E}^4\nonumber
\\& &-36\beta \gamma\omega_{E}+
36\beta b\gamma\omega_{E}+18\beta
\gamma\omega_{E}^2+54\gamma\omega_{E}^3)^\frac{1}{2}),
\end{eqnarray}
\begin{eqnarray}
\lambda^{II}_{2}&=&\frac{1}{4\beta}(6\beta+2\beta^2+18\beta
b+9\beta\omega_{E}+9\omega_{E}^2-2\beta \gamma
+3\gamma\omega_{E}\nonumber
\\& &+(36\beta^2-72\beta^2b+36\beta^2b^2-36\beta^2\omega_{E}
+36\beta^2b\omega_{E}\nonumber
\\& &-108\beta\omega_{E}^2+18\beta^2\omega_{E}^2+216\beta b\omega_{E}^2
+108\beta\omega_{E}^3+162\omega_{E}^4\nonumber
\\& &-36\beta \gamma\omega_{E}+
36\beta b\gamma\omega_{E}+18\beta
\gamma\omega_{E}^2+54\gamma\omega_{E}^3)^\frac{1}{2}).
\end{eqnarray}

\begin{figure*}[htbp]
\begin{center}
\includegraphics{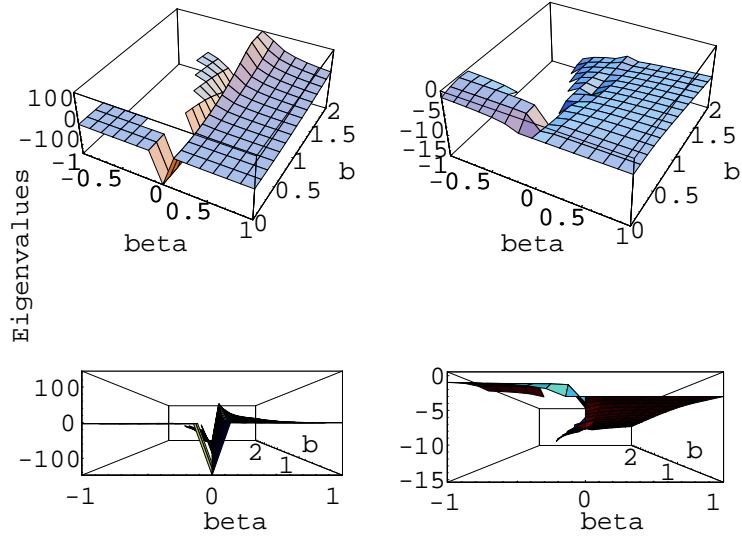}
\end{center}
\caption{Eigenvalues $\lambda^{II}_{1}$(left column) and
$\lambda^{II}_{2}$(right column) of linearizing matrix of the
dynamical system near critical point II for fixed
$\omega_{E}=-1.2$ and $d=2$. The second row is viewed through the
$b$ axial direction on the first row picture. }
\end{figure*}

Because of the complexity of the above two eigenvalue expresses, we
can not simply determine them positive or negative depending on the
parameters. In this paper we mainly perform the numerical
simulations of Eqs.(18) and (19) in  Fig.5. From  Fig.5 we can see
that there are same region of parameters $\beta, b$ where the
eigenvalues $\lambda_{1}$ and $\lambda_{2}$ are both negative. This
means that critical point II is stable point as showed in Fig.1.
Moreover if we investigate the stable point position $(x_{C},
y_{C})$ with the parameters which showed in Figs.2-4, we can find
that the universe will  be dominated by dark matter both the
interaction coefficient $b$  and the  viscous fluid coefficient
$\beta$ become stronger. At the same time, for the critical point
II, we can see that the coordinates $(x_{C}, y_{C})$ in the phase
space is not vanished in its stable region, which tell us that the
coincidence problem will be alleviated in the universe, and we also
can see that the interaction term and the viscous dark energy will
do the same effect on the alleviating the cosmological coincidence
problem.
\section{Conclusions}
In this paper we have investigated the evolution of  the viscous
cosmology model which dark energy interacts with dark matter. Using
the linearizing theory of dynamical system, we found, in our model,
there exists a stable late time scaling solution which corresponds
to the accelerating universe, and we also found the unstable
solution under some appropriate parameters.

We all know that, in order to alleviate the coincidence problem,
some authors considered the effect of quantum correction due to the
conform anomaly: such as dynamical Casimir effect with conformal
anomaly, or dark fluid with conformal anomaly \cite{Brevik}. And
some authors accounted some interacting dark energy models which can
also soften the coincidence problem \cite{wu}.  In this paper that
we found that, if we take into account the bulk viscosity of the
cosmic fluid, the viscosity  softens the coincidence problem as the
interacting dark energy cosmology models. That's to say, both the
non-perfect fluid model and the interacting models of the dark
energy can alleviate or soften the singularity of the universe.
\section{Acknowledgments}
The project is supported by the National Natural Science Foundation
of China under Grant No.10873004, the Scientific Research Fund of
Hunan Provincial Education Department under Grant No. 08B051,
program for excellent talents in Hunan Normal University and the
State Key Development Program for Basic Research Program of China
under Grant No. 2010CB832800
%\newpage

\end{document}